\documentclass[floatfix,aps,amsmath,nofootinbib,twocolumn,10pt]{revtex4}

\usepackage{listings}
\usepackage{graphicx}
\usepackage{bm}
\usepackage{rotating}
\usepackage{array}
\usepackage{amsmath}
\usepackage{amssymb} 
\usepackage{mathrsfs} 
\usepackage{cancel}

\def\({\left(}
\def\){\right)}
\def\[{\left[}
\def\]{\right]}

\def\e{\begin{equation}}
\def\q{\end{equation}}
\def\m{\begin{eqnarray}}
\def\n{\end{eqnarray}}

\begin{document}

\title{Retesting the no-hair theorem with GW150914}

\author{Ke Wang}
\email{wangkey@lzu.edu.cn}
\affiliation{Lanzhou Center for Theoretical Physics, Key Laboratory of Theoretical Physics of Gansu Province,\\
School of Physical Science and Technology, \\
Lanzhou University, Lanzhou 730000, China}
\affiliation{Institute of Theoretical Physics \& Research Center of Gravitation,\\ 
Lanzhou University, Lanzhou 730000, China}

\date{\today}

\begin{abstract}
For a distorted black hole (BH), its ringdown waveform is a superposition of quasi-normal modes (QNMs). In general relativity (GR), the lower order QNM frequencies and damping rates can be well approximated by a polynomial of BH's dimensionless spin and overall scaled by BH's mass. That is to say, we can test the no-hair theorem of BH in GR model-independently by allowing not only an overall fractional deviation (as M. Isi {\it et al.} did) but also a set of fractional deviation for every coefficient. In the paper, we will apply the latter method to retest the no-hair theorem with GW150914 and probe hairs' behaviors if hairs exist. Eventually, we find the data favors GR.

\end{abstract}

\pacs{???}

\maketitle


\section{Introduction}
\label{sec:intro}

The last stage of binary black hole (BBH) coalescence is known as the ringdown of a perturbed black hole (BH). Through perturbation theory, the ringdown waveform can be expressed as a superposition of quasi-normal modes (QNMs)~\cite{Berti:2009kk}. Detecting the loud enough BBH coalescence such as GW150914~\cite{LIGOScientific:2016aoc}, we can directly estimate the QNM frequencies $f_{\ell m n}$ and damping times $\tau_{\ell m n}$ (or rates $\gamma_{\ell m n}\equiv 1/\tau_{\ell m n}$). According to the no-hair theorem of general relativistic BHs, the accurate identification of these QNMs can exclusively determine the mass $M$ and dimensionless spin $\chi$ of a distorted Kerr BH. There are sevaral equivalent sets of fitting formulas between $\{f_{\ell m n}, \tau_{\ell m n}\}$ and $\{M, \chi\}$~\cite{Berti:2005ys,Nagar:2018zoe,London:2018nxs,Isi:2021iql}. While $M$ simply serves as an overall scale on $\{f_{\ell m n}, \tau_{\ell m n}\}$, a polynomial of $\chi$ is need to approximate $\{f_{\ell m n}, \tau_{\ell m n}\}$~\cite{Isi:2021iql}, for example. 

Without a priori assumption that general relativity (GR) is the correct theory (or the remnant BH is the Kerr BH), however, above fitting formulas don't work. More precisely, a reliable detection of QNMs still can't give the information of remnant BH without above priori assumption. The simplest way to deal with this issue is to generalize the fitting formulas and introduce some ad hoc parameters (or hairs) to account for the deviations of remnant BH from Kerr BH in GR~\cite{Isi:2019aib,Carullo:2019flw}. This method enable us to test GR in the strong-field regime (or probe the surrounding fields around a Kerr BH) model-independently. Recently M. Isi {\it et al.} have tested the no-hair theorem with GW150914~\cite{Isi:2019aib} by replacement $f_{221}^{{\rm GR},{\rm Kerr}}\rightarrow f_{221}^{{\rm GR},{\rm Kerr}}(1+\delta f_{221})$ and $\tau_{221}^{{\rm GR},{\rm Kerr}}\rightarrow \tau_{221}^{{\rm GR},{\rm Kerr}}(1+\delta \tau_{221})$. We can easily find that $\delta f_{221}$ and $\delta \tau_{221}$ just overall scale $f_{221}^{{\rm GR},{\rm Kerr}}$ and $\tau_{221}^{{\rm GR},{\rm Kerr}}$. This is to say, there are other ways to introduce some ad hoc hairs. In this paper, we will generalize the fitting formulas \cite{Isi:2021iql} by allowing for fractional deviations from the Kerr BH values in GR for all polynomial coefficients $c_i^{{\rm GR},{\rm Kerr}}$ in turn, as~\cite{Maselli:2019mjd,Carullo:2021dui} did for the parametrized ringdown spin expansion coefficients formalism and as~\cite{Meidam:2017dgf,LIGOScientific:2019fpa,LIGOScientific:2020tif} did for the frequency-domain inspiral-merger-ringdown waveform.

Since both of $M$ and the GR-violating parameters $\{\delta f_{221}, \delta \tau_{221}\}$ of M. Isi {\it et al.}~\cite{Isi:2019aib} serve as an overall scale, there are degeneracies between them. That is to say, the GR-violating parameters $\{\delta f_{221}, \delta \tau_{221}\}$ of M. Isi {\it et al.} contain information about both the source and the underlying theory, which are difficult to disentangle within their approach. In contrast, the hair parameters introduced by our approach are independent of the source by construction and should therefore cleanly encode information only about the underlying theory of gravity. It will become much more useful, especially when we stack data from multiple events.

This paper is organized as follows.
In section~\ref{sec:model}, we give our generalized fitting formulas.
In section~\ref{sec:results}, we test no-hair theorem with GW150914 and give the constraints on the hairs.
Finally, a brief summary and discussion are included in section \ref{sec:sum}.
We adopt geometric units $c=G=1$.
\section{Model}
\label{sec:model}
For a perturbed Kerr BH in GR, its complex-valued ringdown waveform $h\equiv h_+-ih_\times$ is
\begin{equation}
\label{eq:h}
h=\sum_{\ell m n}\mathcal{A}_{\ell m n}~_{-2}S_{\ell m n}(\iota,\varphi)e^{i(t-t_{\ell m n})\tilde{\omega}_{\ell m n}+\phi_{\ell m n}},
\end{equation}
where $\ell>2$ is the azimuthal number, $-\ell \leq m \leq \ell$ is the magnetic number, $0\leq n$ is the overtone number, $\mathcal{A}_{\ell m n}$ are the amplitudes of different modes, $\phi_{\ell m n}$ are the phases of different modes, the spin-weighted spheroidal harmonics $_{-2}S_{\ell m n}(\iota,\varphi)$ are functions of polar angle $\iota$ and azimuthal angle $\varphi$ relative to the BH spin direction~\cite{spin}, $\tilde{\omega}_{\ell m n}=\omega_{\ell m n}+i/\tau_{\ell m n}$ are the complex ringdown frequencies of different modes, $t_{\ell m n}$ are the start times of different modes. Usually $\{\mathcal{A}_{\ell m n}, \phi_{\ell m n}\}$ encode the degree to which each mode is excited during the perturbation of Kerr BH. We can't predict them from perturbation theory and will directly estimate these two nuisance parameters from data. On the contrary, frequencies and damping times of different modes $\{\omega_{\ell m n},\tau_{\ell m n}\}$ exclusively encode the information of mass and dimensionless spin $\{M,\chi\}$ of a perturbed Kerr BH in GR, as is given by fitting formulas~\cite{Berti:2005ys} 
\begin{eqnarray}
2\pi Mf_{\ell m n}=f_{1,\ell m n}+f_{2,\ell m n}(1-\chi)^{f_{3,\ell m n}},\\
\pi f_{\ell m n}\tau_{\ell m n}=q_{1,\ell m n}+q_{2,\ell m n}(1-\chi)^{q_{3,\ell m n}},
\end{eqnarray}
where $2\pi f_{\ell m n}=\omega_{\ell m n}$, the fitting coefficients $f_{i,2 m n}$ and $q_{i,2 m n}$ of lower order modes are listed in Table VIII of~\cite{Berti:2005ys}. There are also fitting formulas in the form of polynomial of $\chi$~\cite{Isi:2021iql}
\begin{eqnarray}
Mf_{\ell m n}=c_{l,\ell m n}\log(1-\chi)+\sum_{i=0}^4c_{i,\ell m n}\chi^i,\\
M\gamma_{\ell m n}=d_{l,\ell m n}\log(1-\chi)+\sum_{i=0}^4d_{i,\ell m n}\chi^i,
\end{eqnarray}
where $\gamma_{\ell m n} = 1/\tau_{\ell m n}$ are the damping rates of different modes, the fitting coefficients $c_{l,i,2 m n}$ and $d_{l,i,2 m n}$ of lower order modes can be computed by the QNM package~\cite{Stein:2019mop}.

For a perturbed non-Kerr BH in modified gravity or a perturbed Kerr BH surrounded by fields in GR, its ringdown waveform still can be described by Eq.~(\ref{eq:h}), the most generic template. Therefore, without the assumption that the BH in question is a perturbed Kerr BH in GR, even though we can estimate $\{\omega_{\ell m n},\tau_{\ell m n}\}$ from data, we can't derive the accurate and complete information of BH. It's quite straightforward to account for the uncertainty by introducing some additional hairs. We generalize the second set of fitting formulas as 
\begin{eqnarray}
\nonumber
\label{eq:f}
&&Mf_{\ell m n}(1+\delta f_{\ell m n})\\
\nonumber
&&=c_{l,\ell m n}(1+\delta c_{l,\ell m n})\log(1-\chi)\\
&&+\sum_{i=0}^4c_{i,\ell m n}(1+\delta c_{i,\ell m n})\chi^i,\\
\nonumber
\label{eq:g}
&&M\gamma_{\ell m n}(1+\delta \gamma_{\ell m n})\\
\nonumber
&&=d_{l,\ell m n}(1+\delta d_{l,\ell m n})\log(1-\chi)\\
&&+\sum_{i=0}^4d_{i,\ell m n}(1+\delta d_{i,\ell m n})\chi^i.
\end{eqnarray}
Recently M. Isi {\it et al.} have tested the no-hair theorem with only $\{\delta f_{221}, \delta \gamma_{221}\}$ as free parameters with GW150914~\cite{Isi:2019aib}. This is also equivalent to independently measuring the frequencies and damping times of $\{220,221\}$ modes and checking their consistency with the assumption of Kerr BH in GR. We can find that there are twelve additional potential hairs $\{\delta c_{l,i,221},\delta d_{l,i,221}\}$ even only for $\{221\}$ mode. Unlike the hairs $\{\delta f_{221}, \delta \gamma_{221}\}$ just overall scale $\{f_{221}, \gamma_{221}\}$, these twelve additional potential hairs $\{\delta c_{l,i,221},\delta d_{l,i,221}\}$ have their own unique dependency on $\chi$. Treating them as free parameters in turn, we can not only test no-hair theorem but also probe hairs' behaviors if hairs exist.
\section{Results}
\label{sec:results} 
M. Isi {\it et al.} have identified multiple ringdown modes $\{220, 221\}$ from the loudest event GW150914~\cite{Isi:2019aib}. Therefore, we download the 4KHz time series of the calibrated GW150914 observable from~\cite{GWOCS} and use this event to estimate hairs $\{\delta c_{l,i,221},\delta d_{l,i,221}\}$ as wells as the other intrinsic parameters $\{\mathcal{A}_{220,221}, \phi_{220,221}, t_{220,221}, M, \chi, \iota,\varphi\}$. Our priors are uniformly distributed: $\delta c_{l,i,221}\in[-10,10]$, $\delta d_{l,i,221}\in[-10,10]$, $\mathcal{A}_{220,221}\in[0,5\times10^{-19}]$, $\phi_{220,221}\in[0,2\pi]$, $t_{220,221}=1126259462.423~{\rm GPS}$, $M\in[50,100]M_{\odot}$, $\chi\in[0,1]$, $\cos\iota\in[-1,1]$ and $\varphi\in[0,2\pi]$. As for the sky position, we fix them as the values consistent with \cite{LIGOScientific:2016vlm}: the right ascension $\alpha=1.95$, the declination $\delta=-1.27~\rm rad$ and the polarization angle $\psi=0.82$. We use the waveform in Eq.~(\ref{eq:h}) accompanied by the generalized fitting formulas in Eq.~(\ref{eq:f}) and Eq.~(\ref{eq:g}) to carry out a Bayesian analysis of GW150914 data. To exclude some unreasonable results, we impose some constraints: $\gamma_{221} > \gamma_{220}>0$, $f_{221} >0$ and $f_{220}>0$. In this paper, we turn to $\mathsf{pyRing}$~\cite{Carullo:2019flw} where the likelihood are defined in the time domain.

\begin{figure*}[]
\begin{center}
\includegraphics[width= 8cm]{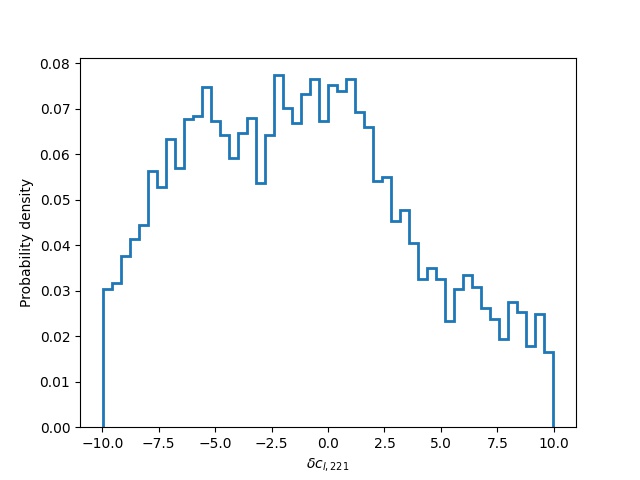}
\includegraphics[width= 8cm]{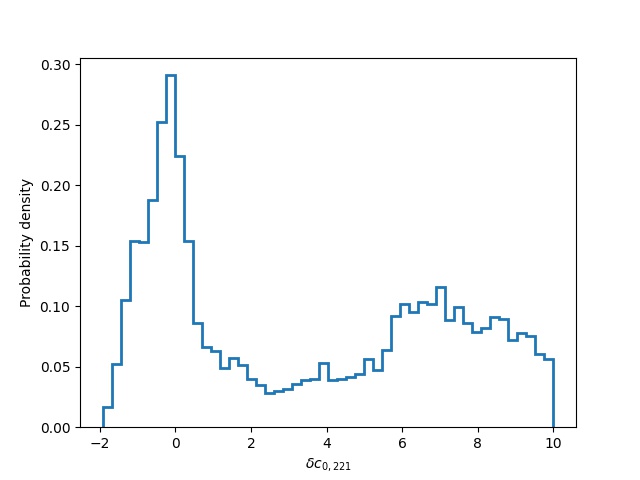}
\includegraphics[width= 8cm]{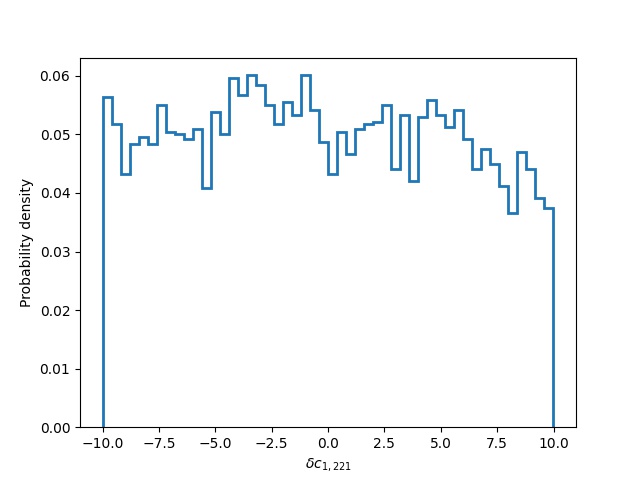}
\includegraphics[width= 8cm]{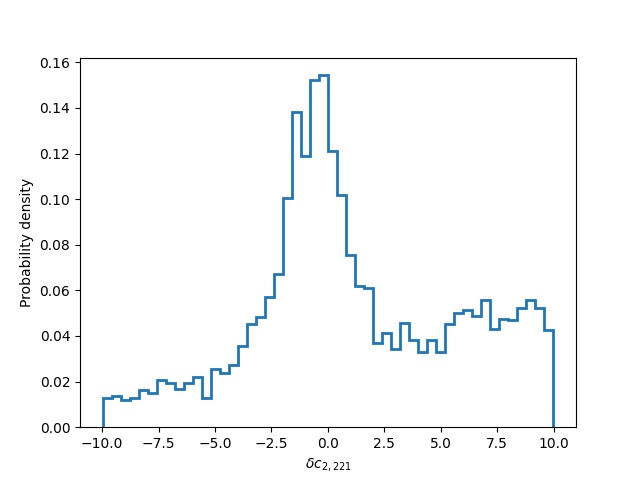}
\includegraphics[width= 8cm]{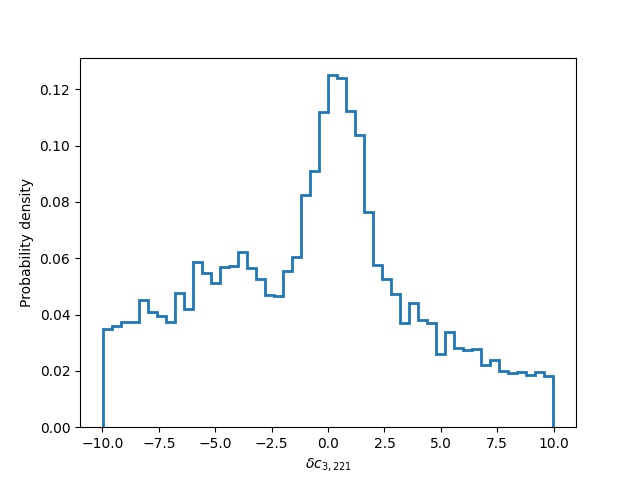}
\includegraphics[width= 8cm]{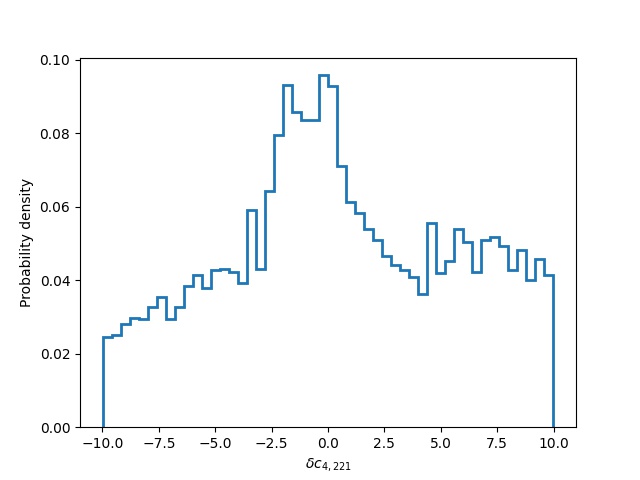}
\end{center}
\caption{Posterior distributions for the set of hairs $\delta c_{l,i,221}$. }
\label{fig:c}
\end{figure*}

Since there are no deviation in the fitting formula for the fundamental $\{220\}$ mode, the long live part of the whole data can constrain $\{M,\chi\}$. Therefore, the individual degeneracy between $\{\delta c_{l,i,221},\delta d_{l,i,221}\}$ and $\chi$ is broken slightly. Fig.~\ref{fig:c} shows the constraints on one of the six $\delta c_{l,i,221}$ individually, keeping the other deviation parameters fixed to its GR value of $0$. 
For $\delta c_{1,221}$, we obtain an almost flat posterior distribution.
And for the other hairs $\{\delta c_{l,i\neq1,221}\}$, there are obvious peaks around the no-hair hypothesis $\delta c_{l,i\neq1,221}=0$ in their posterior distribution.
Also we summarize the log Bayes factors for every model with one possible hair $\delta c_{l,i,221}$ versus the model in GR in Table~I, from which we find that the data also favors GR.

\begin{figure*}[]
\begin{center}
\includegraphics[width= 8cm]{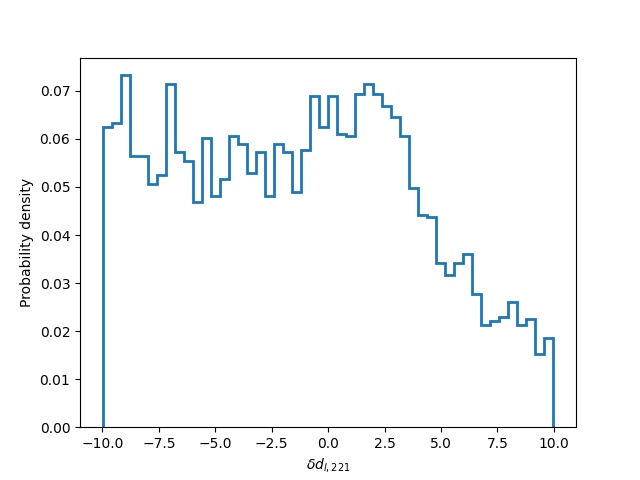}
\includegraphics[width= 8cm]{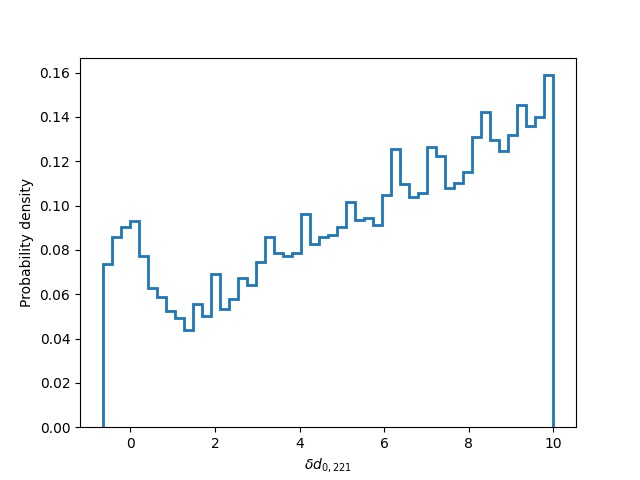}
\includegraphics[width= 8cm]{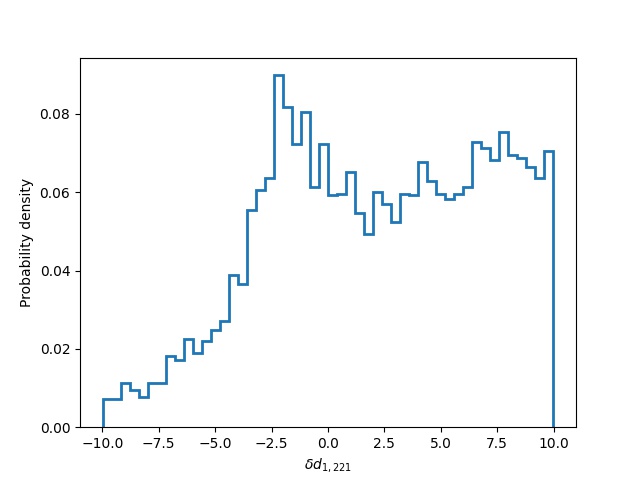}
\includegraphics[width= 8cm]{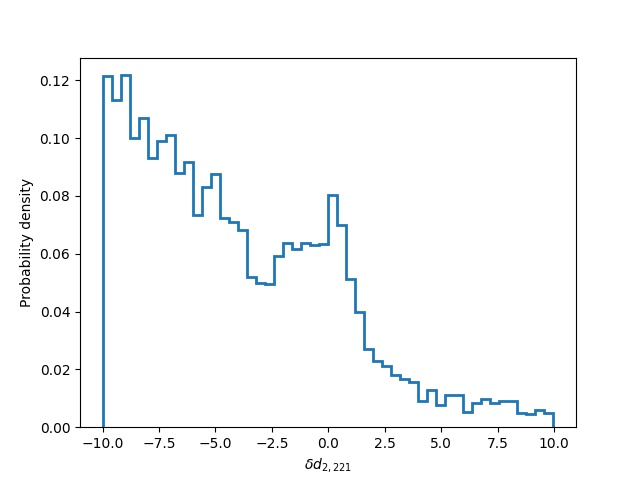}
\includegraphics[width= 8cm]{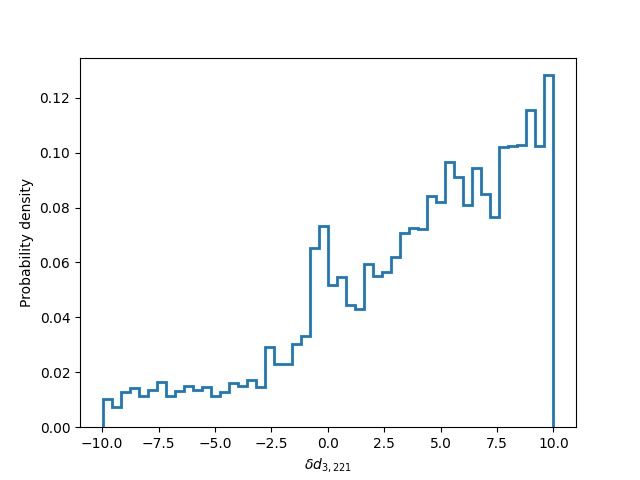}
\includegraphics[width= 8cm]{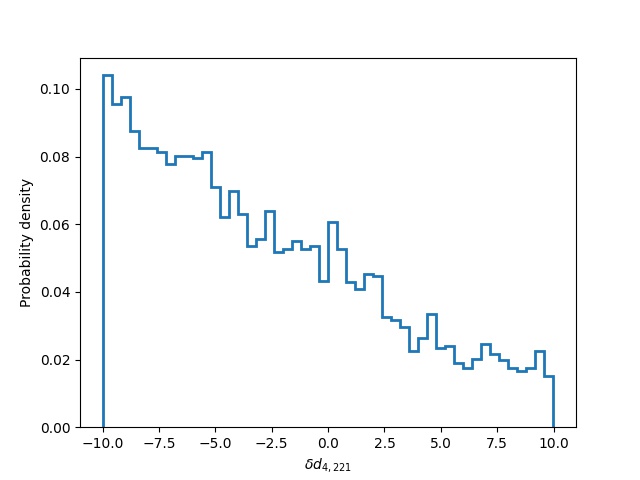}
\end{center}
\caption{Posterior distributions for the set of hairs $\delta d_{l,i,221}$.}
\label{fig:d}
\end{figure*}

Similarly, Fig.~\ref{fig:d} shows the constraints on one of the six $\delta d_{l,i,221}$ individually, keeping the other deviation parameters fixed to its GR value of $0$. 
The direction of the tails of the distribution is determined by the coefficients $d_{l,i,221}$. 
Given $d_{2,221}<0$, for example, the upper limit of $\delta d_{2,221}$ (or the lower limit of $\gamma_{221}$) is well constrained under the constraints $\gamma_{221} > \gamma_{220}>0$.
On the hand, the data prefers a smaller $\delta d_{2,221}$ (or a larger $\gamma_{221}$). However, a large enough $\gamma_{221}$ is equivalent to no overtone $\{221\}$ instead of an obvious hair $\delta d_{2,221}$.
Therefore, there is still no evidence that hairs $\delta d_{l,i,221}$ exist according to Fig.~\ref{fig:d} and the $\log \rm {B^{Hair}_{GR}}$ for $\delta d_{l,i,221}$ as listed in Table~I.

\begin{table*}
\label{tab:B}
\caption{Summary of the log Bayes factors for one possible hair versus GR.}
\begin{tabular}{p{1.1cm} <{\centering}p{1.1cm}<{\centering} p{1.1cm}<{\centering}p{1.1cm}<{\centering}p{1.1cm}<{\centering}p{1.1cm}<{\centering}p{1.1cm}<{\centering}<{\centering}p{1.1cm}<{\centering} p{1.1cm}<{\centering}p{1.1cm}<{\centering}p{1.1cm}<{\centering}p{1.1cm}<{\centering}p{1.1cm}<{\centering}   }
\hline
Hair & $\delta c_{l,221}$ & $\delta c_{0,221}$ & $\delta c_{1,221}$ & $\delta c_{2,221}$ & $\delta c_{3,221}$ & $\delta c_{4,221}$ & $\delta d_{l,221}$ & $\delta d_{0,221}$ & $\delta d_{1,221}$ & $\delta d_{2,221}$ & $\delta d_{3,221}$ & $\delta d_{4,221}$  \\
\hline
$\log \rm {B^{Hair}_{GR}}$ & $-0.83$ & $-1.60$ & $0.02$ & $-1.33$ & $-1.37$ & $-0.83$ & $-0.48$ & $-0.73$ & $-0.34$ & $-0.76$ & $-0.48$ & $-0.30$ \\
\hline
\end{tabular}
\end{table*}

\section{Summary and discussion}
\label{sec:sum} 
We generalize the fitting formulas between $\{f_{221}, \tau_{221}\}$ and $\{M, \chi\}$ by introducing twelve additional potential hairs $\{\delta c_{l,i,221},\delta d_{l,i,221}\}$. Then we use modified $\mathsf{pyRing}$~\cite{Carullo:2019flw} to constrain these deviation parameters. 
From the posterior distribution and the log Bayes factors for $\{\delta c_{l,i,221}\}$, we find the data favors GR. 
The case for $\{\delta d_{l,i,221}\}$ is complicated. The data prefers additional hairs $\{\delta d_{l,i,221}\}$ which leads to a larger $\gamma_{221}$. However, a large enough $\gamma_{221}$ is equivalent to no overtone $\{221\}$. Meanwhile, the log Bayes factors for $\{\delta d_{l,i,221}\}$ also favors GR.

Also we can set $t_{220,221}=1126259462.423+0.003~{\rm GPS}$ to ignore the overtone, $\{221\}$ mode~\cite{Carullo:2019flw}. In this case, there should be another twelve deviation parameters $\{\delta c_{l,i,220},\delta d_{l,i,220}\}$ instead of $\{\delta c_{l,i,221},\delta d_{l,i,221}\}$. And the constraint on $\{M,\chi\}$ will correlate with the constraint on $\{\delta c_{l,i,220},\delta d_{l,i,220}\}$, hence the posteriors for them will broaden.

It is worth noting that there may be some hairs beyond the fourth order $\chi^4$ even though their coefficients are zero in no-hair case. However, these more general case is out of our reach due to the lack of theoretical hints and the increase of parameter.

~

~

~

\vspace{5mm}
\noindent {\bf Acknowledgments}
We acknowledge the use of HPC Cluster of Tianhe II in National Supercomputing Center in Guangzhou. Ke Wang is supported by grants from NSFC (grant No. 12005084 and grant No. 12047501).



\end{document}